\documentclass{emulateapj}  
\slugcomment{Submitted to \aj}

\shortauthors{Auger et al.}
\shorttitle{Gravitational Lens - Galaxy Group Connection II}

\begin{document}

\title{The Gravitational Lens -- Galaxy Group Connection. II. Groups Associated with B2319+051 and B1600+434}

\author{M. W. Auger, C. D. Fassnacht, A. L. Abrahamse, L. M. Lubin}
\affil{
   Department of Physics, University of California, 1 Shields Avenue,
   Davis, CA 95616 }
\email{mauger@physics.ucdavis.edu, fassnacht@physics.ucdavis.edu}

\author{G. K. Squires}
\affil{
   Spitzer Science Center, California Institute of Technology,
   Mail Code 220-6, 1200 E. California Blvd., Pasadena, CA 91125 }

\begin{abstract}
We report on the results of a spectroscopic survey of the environments of the gravitational lens systems CLASS B1600+434 ($z_l = 0.41, z_s = 1.59$) and CLASS B2319+051 ($z_l = 0.62$). The B1600+434 system has a time delay measured for it, and we find the system to lie in a group with a velocity dispersion of $100~{\rm km}\,{\rm s}^{-1}$ and at least six members.  B2319+051 has a large group in its immediate foreground with at least 10 members and a velocity dispersion of $460~{\rm km}\,{\rm s}^{-1}$ and another in the background of the lens with a velocity dispersion of $190~{\rm km}\,{\rm s}^{-1}$.  There are several other small groups in the fields of these lens systems, and we describe the properties of these moderate redshift groups. Furthermore, we quantify the effects of these group structures on the gravitational lenses and find a $\sim 5\%$ correction to the derived value of $H_0$ for B1600+434. 
\end{abstract}


\section{INTRODUCTION}

Galaxy groups are a fundamental component of the Large Scale Structure of the Universe and over half of all galaxies in the local Universe are members of groups \citep[e.g.,][]{tully,ramella}.  Precision cosmology relies on an accurate knowledge of the distribution of galaxies in order to model the galaxy power spectrum properly \citep[e.g.,][]{evrard,yang}, and it is essential that galaxy groups are identified to describe the small-scale, non-linear regime of the matter power spectrum accurately.  Several local redshift surveys, including the Sloan Digital Sky Survey \cite[SDSS,][]{york} and the 2 Degree Field Galaxy Redshift Survey \cite[2dFGRS,][]{colless}, have produced results using groups out to $z \sim 0.2$ \citep{weinmann,collister,abazajian,padilla}. At higher redshifts, the Canadian Network for Observational Cosmology redshift survey \citep[CNOC2,][]{carlberga} probes redshifts $0.15 \lesssim z \lesssim 0.5$, while the DEEP2 Galaxy Redshift Survey is probing redshifts from $0.7 \lesssim z \lesssim 1$ \citep{coil2006,coil2004}.  Weak lensing analyses of groups also promise to yield information about the mass distribution in galaxy groups \citep[e.g.,][]{moller,faure,hoekstra}.  Knowledge of group substructure can then be used to confirm or refute dark matter models \citep{reed,donghia}.

Groups also hold significance for strong gravitational lensing. \citet{keeton2000} estimate that 25\% of lenses lie in group environments, though the fraction could be much higher \citep[e.g.,][]{williams,blandford,oguri}. A moderately massive group that is sufficiently close to the lensing galaxy can contribute to the convergence in the lensing potential and alter the shear field of the lens. Theoretical estimates of the importance of these contributions have been made \citep[e.g.,][]{oguri,keeton2004,moller} but it remains unclear to what extent lens models must account for group environments \citep[e.g.,][]{momcheva,fassnacht2005,morgan,dalal}. Furthermore, gravitational lens time-delay measurements and improved mass modeling techniques have lowered the uncertainties in measurements of the Hubble Constant, $H_0$, to the $\approx$ 10\% level \citep[e.g.,][]{koopmans2003,kochanek2004}, and the systematic errors introduced by ignoring lens environments now contribute significantly to the $H_0$ error budget.

In addition to improving the determination of cosmological parameters, galaxy groups also allow us to study the evolution of galaxies.  Group environments are likely locations for mergers \citep[e.g.,][]{zabludoff,carlbergb,aarseth} compared to clusters or the field \citep[e.g.,][]{aceves,lin,conselice,patton}.  Mergers and non-merger interactions in groups can drive changes in galaxy morphologies and star formation rates (SFR) that are suppressed in more dense environments \citep[e.g.,][]{zabludoff,bower}.  Furthermore, active galactic nuclei (AGN) are expected to reside in dense environments where galaxy-galaxy interactions provide mechanisms for fueling AGN \citep[e.g.,][]{best2004,bahcall}, though the validity of this claim is uncertain \citep[e.g.,][]{kauffmann,miller}.  \cite{best2005} only find a correlation between \emph{radio-loud} AGN and the local environment, although \citet{mclure} find no distinction between radio-loud and radio-quiet AGN populations in clusters.

We are currently conducting a systematic survey to investigate the environments of strong gravitational lens systems \citep[e.g.,][]{fassnacht2002,fassnacht2005}. In this paper we present the discovery of three groups associated with the lens systems B1600+434 \citep{jackson} and B2319+051 \citep{rusin}, hereafter B1600 and B2319. These lenses were discovered by the Cosmic Lens All-Sky Survey \citep[CLASS;][]{myers,browne}.  In the case of B2319, the lens is not found to be part of a group, but there is a large group in the immediate foreground of the lens and another group in the background of the lens.  In contrast, B1600 is found to be a member of a poor group of galaxies. Here we examine the effects of these new groups on the lens models for each system, quantifying the contributions of the environments to the shears and convergences of the lens systems. The final sample of groups from our study will be compared with groups observed in the local Universe \citep{merchan,balogh} and with recent observations of other moderate redshift groups \citep{wilmana,wilmanb,momcheva,williams,gerke} to quantify the evolution of groups with redshift.

\section{OBSERVATIONS AND DATA REDUCTION}
There are approximately 80 known strong gravitational lens systems, and many of these require only very small \emph{ad hoc} additional shear and convergence components to be adequately modeled \cite[e.g.,][]{lehar}. In contrast, the B2319 system ($z_l = 0.62$) requires a large shear component to fit acceptable models, and this shear cannot be entirely accounted for by the lensing galaxy \citep{rusin}.  There is a secondary lensing candidate 3\farcs4 away from the primary lens, and the redshifts of the lensing and secondary galaxies are 0.624 and 0.588, respectively \citep{lubin}.  However, it is likely that the secondary galaxy alone cannot account for the shear discrepancy, and a more massive, group-like structure is required to reproduce the observed image configuration.  The B1600 system ($z_l = 0.41, z_s = 1.59$) has a reasonably well constrained lens model \citep{koopmans1998,maller}, but initial spectroscopy of the environment of the lensing galaxy has revealed several nearby galaxies with redshifts similar to the lensing galaxy. It is crucial that the group associated with B1600 be investigated due to the importance of this system in determining $H_0$ \citep{koopmans2000,maller,kochanek2002,kochanek2003,burud}.

\subsection{Imaging}
We have obtained deep non-photometric BVRI images of B2319 using the Low-Resolution Imaging Spectrometer \citep[LRIS;][]{oke} and Echellette Spectrograph and Imager \citep[ESI;][]{sheinis} instruments on the Keck Telescopes.  We have also obtained photometric LRIS snapshots of the field in the BVI filters, allowing us to photometrically calibrate our deeper imaging in these bands.  These Keck imaging data were reduced using standard IRAF\footnote{IRAF is distributed by the National Optical Astronomy Observatories, which are operated by the Association of Universities for Research in Astronomy, Inc., under cooperative agreement with the National Science Foundation.} tasks.  Our primary imaging of B1600 comes from deep BRI imaging from the Suprime-Cam instrument \citep{miyazaki} on the Subaru Telescope, obtained from the SMOKA archive \citep{baba}.  These data were reduced using the SDFRED package \citep{ouchi}.  A comparison between stars in our Suprime-Cam imaging and SDSS imaging of the field of B1600 (after applying Lupton's transformation\footnote{We used the filter transformations that had smaller reported values for sigma.  Lupton's filter transformation equations can be found on the SDSS Data Release 4 website at: http://www.sdss.org/dr4/algorithms/sdssUBVRITransform.html} between the two different filter sets) allowed us to establish a photometric zeropoint for the Suprime-Cam archival data.  Additionally, we have \emph{Hubble Space Telescope} (\emph{HST}) imaging of both fields with the WFPC2 camera in the F555W and F814W filters.  A summary of our imaging data can be found in Table \ref{table_imaging}.

\subsection{Spectroscopy}
The candidate group members for the two lens systems were chosen based upon the colors of the lensing galaxy or other galaxies in the field at the same redshift as the lens. Target galaxies generally had colors within 0.1~mag of the colors of the lensing galaxy and were within 200\arcsec of the lensing galaxy. Four slitmasks were taken for the field of B2319 and one slitmask was obtained for the field of B1600 with LRIS (Table \ref{table_spectra}).  A handful of additional redshifts were also obtained from longslit spectra of each lens system with LRIS and ESI.  The spectra were reduced using standard IRAF tasks, and redshifts were determined by finding at least one emission line and one other feature, or by identifying multiple absorption features in each spectrum. Redshift errors are typically $\Delta z \approx 0.0004$. In total, we have 53 redshifts for the B2319 field and 24 redshifts for the B1600 field (the number of redshifts in Table \ref{table_spectra} does not reflect our longslit spectra but do include multiple observations of some target galaxies; all five targets with repeated observations have redshifts that agree within the measurement errors).

\begin{deluxetable*}{lcrcr}
\tabletypesize{\scriptsize}
\tablecolumns{5}
\tablewidth{0pc}
\tablecaption{Lens System Imaging}
\tablehead{
 \colhead{} &
 \colhead{} &
 \colhead{Exposure} &
 \colhead{} &
 \colhead{} \\
 \colhead{Lens System} &
 \colhead{Band} &
 \colhead{Time (s)} &
 \colhead{Instrument} &
 \colhead{Date}
}
\startdata
B2319+051 & R & 1200 & LRIS & 1998 Aug 01 \\
 & B & 900 & LRIS & 1999 Aug 16 \\
 & V & 3900 & LRIS & 1999 Aug 16 \\
 & I & 2700 & LRIS & 1999 Aug 16 \\
 & F555W & 4800 & WFPC2 & 2000 Sep 27 \\
 & F814W & 4800 & WFPC2 & 2000 Sep 27 \\
B1600+434 & B & 1200 & Suprime-Cam & 2001 May 21 \\
 & R & 2700 & Suprime-Cam & 2001 Apr 25,26 \\
 & I & 1380 & Suprime-Cam & 2001 Apr 20 \\
 & F555W & 5300 & WFPC2 & 2001 Jun 16 \\
 & F555W & 18200 & WFPC2 & 2001 Sep 21-23 \\
 & F814W & 7800 & WFPC2 & 2001 Sep 26 \\
\enddata
\label{table_imaging}
\end{deluxetable*}

\begin{deluxetable}{lccr}
\tabletypesize{\scriptsize}
\tablecolumns{4}
\tablewidth{0pc}
\tablecaption{Lens Field LRIS Spectroscopy}
\tablehead{
\colhead{} &
\colhead{Number of} &
\colhead{Exposure} &
\colhead{} \\
\colhead{Lens System}
 & \colhead{Redshifts}
 & \colhead{Time (s)}
 & \colhead{Date}
}
\startdata
B2319+051 & 8 & 5400 & 2001 Jul 26 \\
  & 21 & 8100 & 2002 Jul 15 \\
  & 12 & 3600 & 2002 Jul 16 \\
  & 15 & 5400 & 2003 Aug 01 \\ \\
B1600+434 & 21 & 5400 & 2003 Jul 31 \\
\enddata
\label{table_spectra}
\end{deluxetable}

\section{GROUP IDENTIFICATION}
\begin{figure}
\epsscale{1}
\plotone{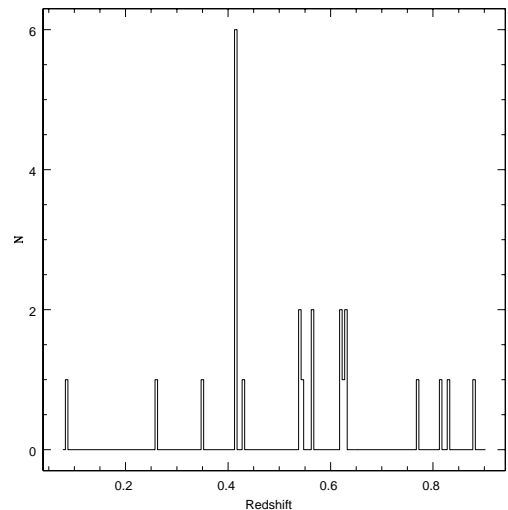}
  \caption{Redshift distribution of galaxies in the field of the gravitational lens B1600+434. The redshift of the lens is 0.41.}
\label{fig_redshift_1600}
\end{figure}

\begin{figure}
\epsscale{1}
\plotone{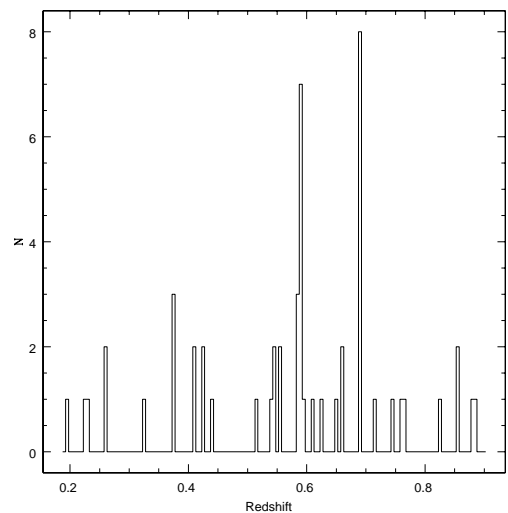}
  \caption{Redshift distribution of field galaxies for the gravitational lens B2319+051. The redshift of the lens is 0.62.}
\label{fig_redshift_2319}
\end{figure}

The redshift distributions obtained from the spectroscopy of our two target lens systems are shown in Figures \ref{fig_redshift_1600} and \ref{fig_redshift_2319}.  There are two obvious peaks in the field of B2319 and one clear peak at the lens redshift of B1600.  We identify possible groups by initially associating all galaxies that are within $\delta z = 0.005$ of each other (in effect, we take each spike from the redshift distribution to be a potential group).  We then use the formalism of \cite{wilmana} to exclude non-members and to determine the group's velocity dispersion.  That is, we find the average redshift and position of the potential group members (clipping the extreme members for groups with more than 3 potential members), define a first approximation observed velocity dispersion of $\sigma_{obs} = 350~(1 + \bar{z})~{\rm km}\,{\rm s}^{-1}$, set an initial redshift shell
$$
\delta z = \frac{2\,\sigma_{obs}}{c},
$$
and specify a maximum angular radius based upon this redshift shell \citep[see][for example]{wilmana}:
$$
\delta\theta = 206265\arcsec \frac{c~\delta z}{b(1 + \bar{z}) H_{0} D_{\theta}} .
$$
Here, $D_{\theta}$ is the angular diameter distance at the mean redshift.  We will assume a $\Lambda$CDM cosmology with $\Omega_{m} = 0.27$ and $\Omega_{\Lambda} = 0.73$, and we follow \cite{wilmana} in fixing our aspect ratio, $b = 3.5$.

Upon excluding potential members outside of our redshift shell $\delta z$ and our maximum radius $\delta\theta$, we compute the velocity dispersion of the remaining group members.  We calculate $\sigma_{obs}$ using the standard deviation, the gapper algorithm, and a clipped standard deviation.  Each of these methods resulted in the same group membership, and we report the results obtained from the gapper algorithm to allow direct comparison with the \cite{wilmana} sample.  We use the updated value for $\sigma_{obs}$ to determine a new $\delta z$ and $\delta\theta$ and repeat the process until no more potential members are eliminated; for each potential group, stable membership was achieved in two iterations.

\begin{figure*}
\begin{center}
\epsscale{0.2}
 \includegraphics[width=0.3\textwidth]{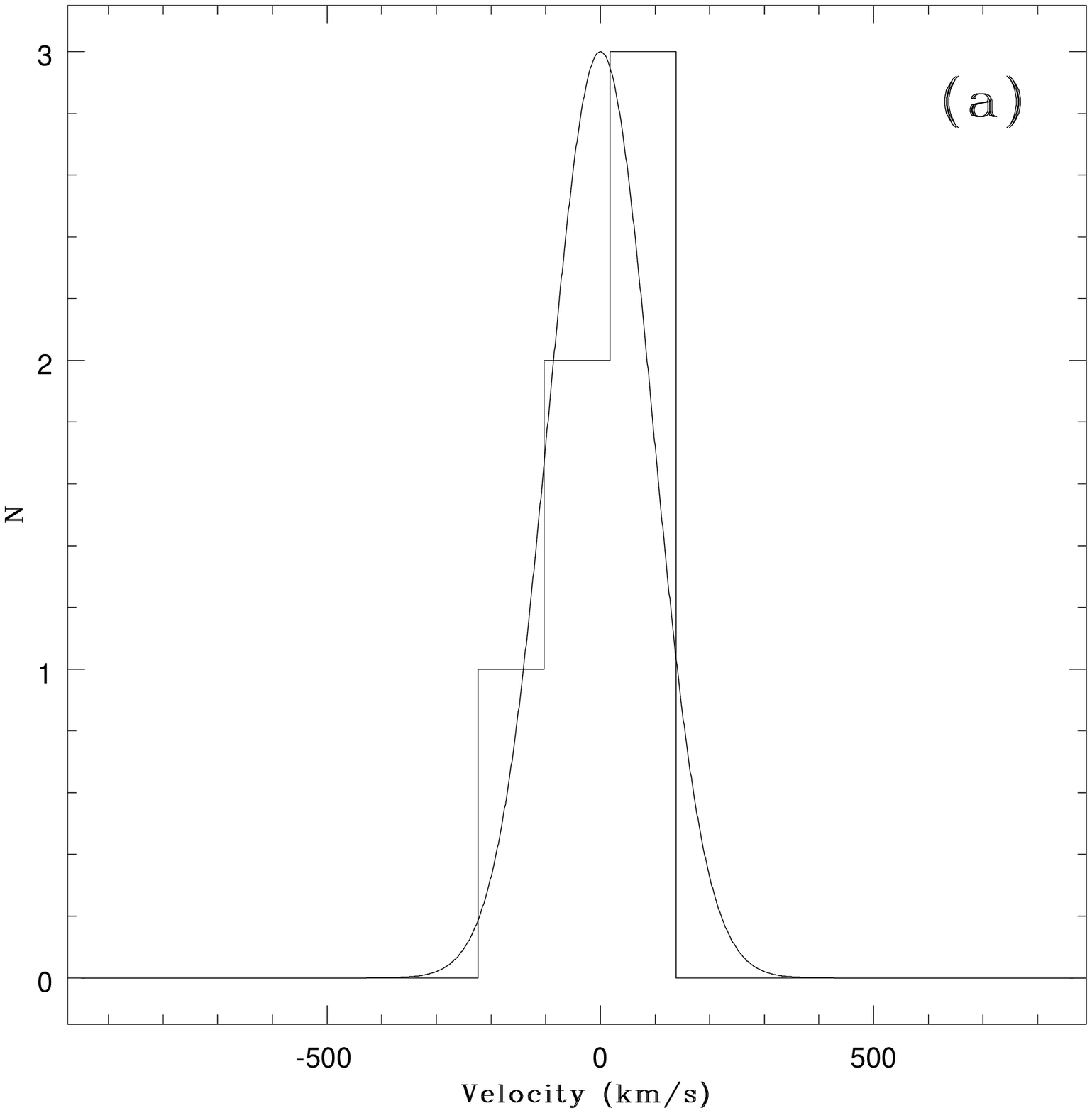}
 \includegraphics[width=0.3\textwidth]{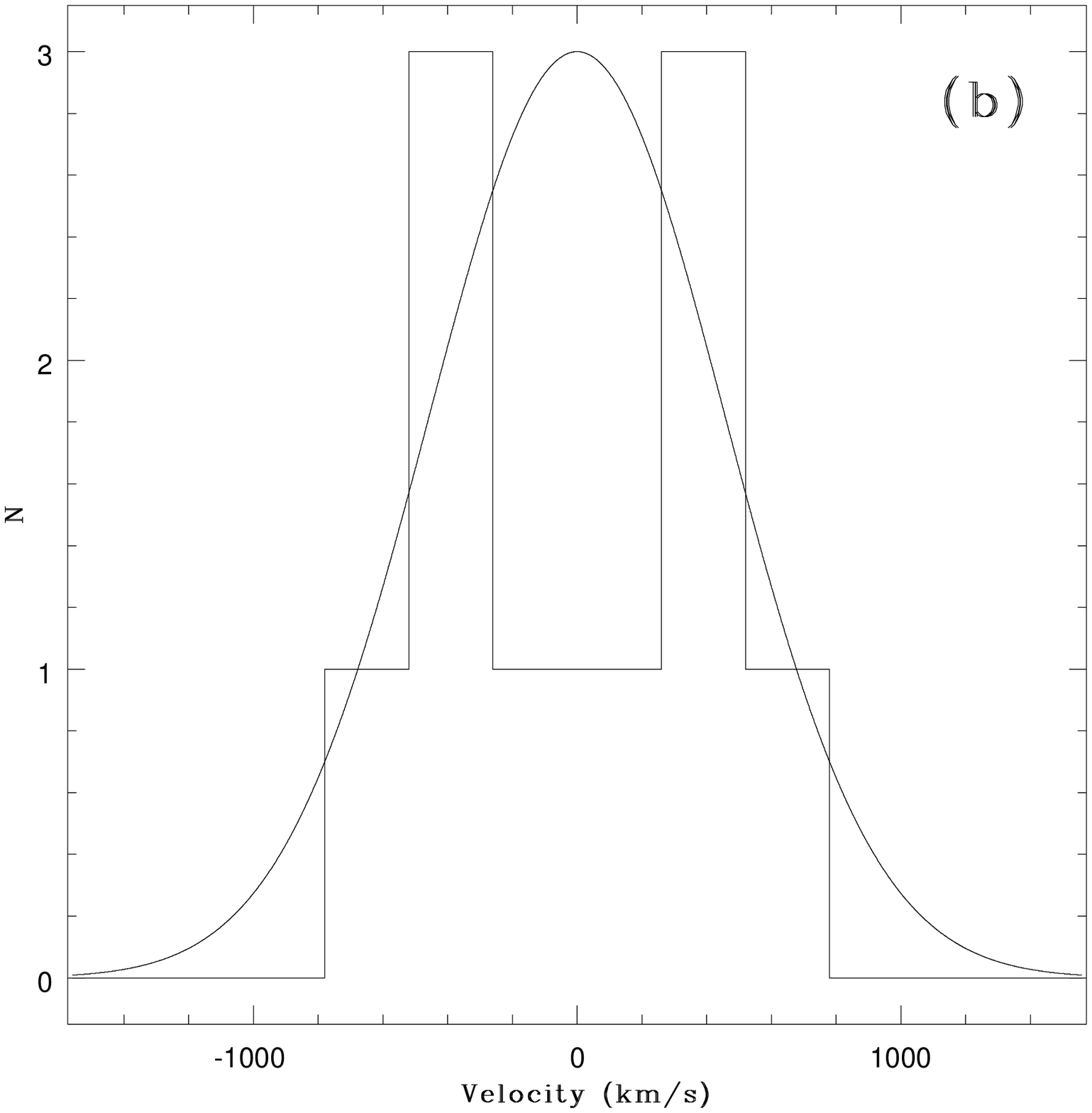}
 \includegraphics[width=0.3\textwidth]{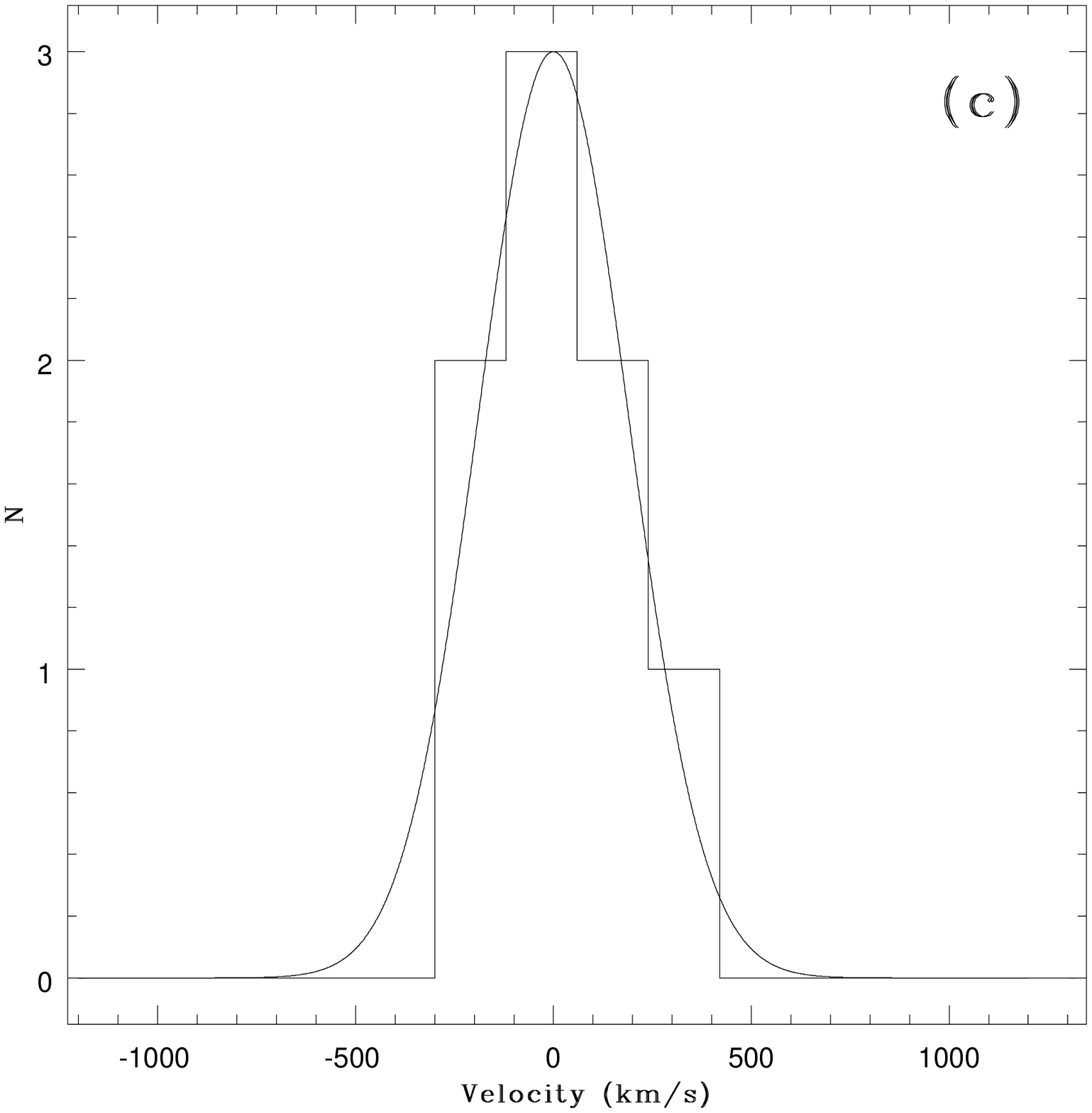}
  \caption{Velocity histograms for members of the groups associated with (a) B1600+434 and the (b) foreground and (c) background groups of B2319+051.}
\label{velocity}
\end{center}
\end{figure*}

We retain any associations with three or more members.  This leaves us with three groups in the field of B1600 and four groups in the field of B2319.  The details for these groups are listed in Table \ref{table_groups} and the velocity distribution of each group containing six or more members is shown in Figure \ref{velocity}. Errors in the velocity dispersion were determined using a jacknife analysis; errors for groups with 3 members are not meaningful. The group located at $z = 0.5894$ in the field of B2319 could be two smaller groups interacting, as indicated by the bimodal velocity distribution, or it could be a small cluster.  Until we obtain more redshift information for the field, we will treat it as a single group.

\section{GROUP PROPERTIES}
As outlined in Table \ref{table_groups}, we have detected three larger groups with six or more members and four smaller groups. Whether many of these smaller groups are bound systems is questionable. For example, the two small groups associated with B1600+434 have significantly larger velocity dispersions than typical small groups \citep[e.g.,][]{merchan}. We report parameters for these small groups, but we limit our discussion to the three groups with more than five members.  The velocity dispersions of these groups are consistent with other groups with comparable numbers of members \citep{wilmana}. In determining group parameters, we have assumed that the groups are relaxed and we use the virial theorem to determine masses and radii.

\begin{deluxetable*}{lcccclcccc}
\tabletypesize{\scriptsize}
\tablecolumns{10}
\tablewidth{0pc}
\tablecaption{Summary of Galaxy Groups in the Fields of B1600 and B2319}
\tablehead{
 &
 &
 &
 &
 &
 \colhead{$\sigma$} &
 \colhead{$M_{vir}$} &
 \colhead{$R_{vir}$} &
 \colhead{$R_{p}$} &
 \colhead{$t_{c}$} \\
 \colhead{Field} &
 \colhead{RA} &
 \colhead{Dec} &
 \colhead{$z$} &
 \colhead{N$_{mem}$} &
 \colhead{(${\rm km}\,{\rm s}^{-1}$)} &
 \colhead{($10^{14}~h~{\rm M}_{\sun}$)} &
 \colhead{($h^{-1}$~Mpc)} &
 \colhead{($h^{-1}$~Mpc)} &
 \colhead{(Hubble times)}
}
\startdata
B1600 & 16 01 40 & 43 16 48 & 0.415 & 6 & $100 \pm 40$ & 0.19 & 0.16 & 0.33 & 0.04 \\
 & 16 01 40 & 43 17 44 & 0.543 & 3 & $640 \pm 770$ & 7.25 & 1.02 & 0.93 & 0.04 \\
 & 16 01 39 & 43 17 09 & 0.629 & 5 & $840 \pm 350$ & 9.6\phn & 1.02 & 0.99 & 0.03 \\
 \\
B2319 & 23 21 39 & 05 26 52 & 0.589 & 10 & $460 \pm 80$ & 0.67 & 0.24 & 0.42 & 0.01 \\
 & 23 21 40 & 05 26 48 & 0.689 & 8 & $190 \pm 50$ & 0.15 & 0.32 & 0.41 & 0.05 \\
 & 23 21 41 & 05 27 43 & 0.375 & 3 & $340 \pm 240$ & 0.64 & 0.43 & 0.62 & 0.03 \\
 & 23 21 40 & 05 26 21 & 0.542 & 3 & $260 \pm 290$ & 1.03 & 1.10 & 1.16 & 0.11 \\
\enddata
\label{table_groups}
\end{deluxetable*}

We are hesitant to quote group properties due to the incompleteness of our group member sampling. Nevertheless, we provide estimates for the virial radius, mean pairwise separation, mass, and crossing time for each group; the values listed in Table \ref{table_groups} for these properties should be considered order of magnitude estimates. Following the definitions of \citet{ramella}, we define the group's virial radius
$$
R_{vir} = \frac{\pi}{2}D_{\theta}N_{pairs}\left(\sum_{i}\sum_{j>i}\theta_{ij}^{-1}\right)^{-1},
$$
the mean (physical) separation between pairs
$$
R_{p} = \frac{4}{\pi}D_{\theta}N^{-1}_{pairs}\sum_{i}\sum_{j>i}\theta_{ij},
$$
the virial mass
$$
M_{vir} = \frac{6\sigma^2R_{vir}}{G},
$$
and the crossing time
$$
t_{c} = \frac{3R_{vir}H_0}{5^{3/2}\sigma}
$$
where $D_{\theta}$ is the angular diameter distance at the group's redshift, $N_{pairs}~=~N_{mem}(N_{mem}-1)/2$ is the number of unique group member pairs, $\theta_{ij}$ is the angular separation between group members, and $t_{c}$ is measured in Hubble times.

\begin{figure*}[ht]
\epsscale{1}
\plotone{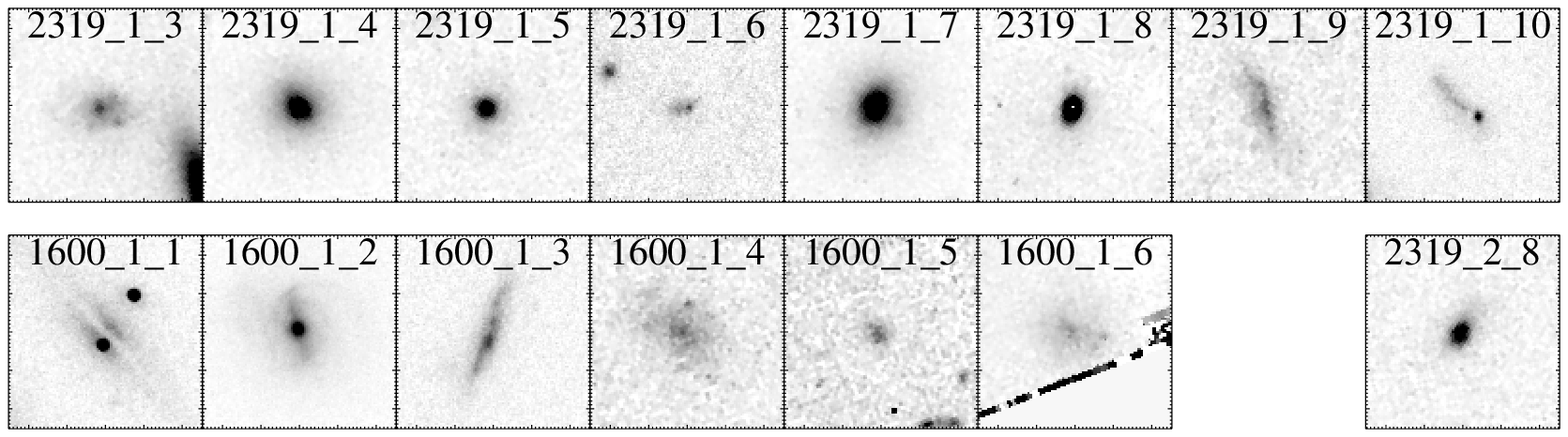}
  \caption{HST WFPC2 imaging of the B1600 and B2319 lens systems; each image is 5\farcs05 on a side. The images for 1600\_1\_4 and 1600\_1\_5 are taken from the F555W filter, and the rest of the images are from the F814W filter. The lens in B1600 is 1600\_1\_1.}
\label{hst_cutout}
\end{figure*}

\emph{HST} imaging is used to determine morphological classifications of group members that lie within the WFPC2 field of view (Figure \ref{hst_cutout}), which includes most galaxies within $\sim 90\arcsec$ of the lensing galaxy.  Additionally, we indicate whether the group galaxies have strong emission lines present in their spectra. Considering the nature of the morphological and spectral features for each galaxy, we characterize the galaxies as either late-(spiral/irregular morphologies or strong emission features and poor morphological information) or early-type galaxies (early-type or uncertain morphologies and no emission features).

We find that all of the galaxies that we associate with B1600 are late-type galaxies (see Table \ref{table_1600}).  With a velocity dispersion of $\approx 100~{\rm km}\,{\rm s}^{-1}$ and only six members, this early-type fraction of zero is similar to some of the results of \cite{zabludoff}, who found that groups with no early type galaxies had lower velocity dispersions and fewer members than groups with non-zero early-type fractions.  \citet{dai} do not find any X-ray emission associated with the B1600 group, which is also expected for groups with low velocity dispersions and low early-type fractions \citep[e.g.,][]{zabludoff,mulchaey}. The absence of X-ray emission indicates that the group may not be relaxed, in which case the mass reported for the group is not valid.

\begin{deluxetable*}{lccccccccc}
\tabletypesize{\scriptsize}
\tablecolumns{9}
\tablewidth{0pc}
\tablecaption{Members of B1600 Group}
\tablehead{
\colhead{Label} &
\colhead{RA} &
\colhead{Dec} &
\colhead{$z$} &
\colhead{B\tablenotemark{a}} &
\colhead{R\tablenotemark{a}} &
\colhead{I\tablenotemark{a}} &
\colhead{Morphology\tablenotemark{b}} &
\colhead{Emission} &
}
\startdata
1600\_1\_1 & 16 01 40.48 & 43 16 48.0 & 0.4144 & 21.50 & 19.91 & 19.22 & Sa & Yes\\
1600\_1\_2 & 16 01 40.84 & 43 16 45.2 & 0.4146 & 21.84 & 19.61 & 18.87 & SBa & Yes\\
1600\_1\_3 & 16 01 39.58 & 43 16 48.3 & 0.4151 & 23.60 & 21.39 & 20.54 & Sc & Yes\\
1600\_1\_4 & 16 01 42.83 & 43 17 01.1 & 0.4140 & 23.60 & 21.96 & 21.53 & Sd & Yes\\
1600\_1\_5 & 16 01 41.83 & 43 18 03.6 & 0.4149 & 25.36 & 24.04 & 23.76 & Irr & Yes\\
1600\_1\_6 & 16 01 36.16 & 43 15 23.2 & 0.4150 & 23.32 & 21.78 & 21.35 & Irr & Yes\\
\enddata
\tablenotetext{a}{From Subura Suprime-Cam imaging}
\tablenotetext{b}{Determined from WFPC2 F555W and F814W imaging}
\label{table_1600}
\end{deluxetable*}

\begin{deluxetable*}{lccccccccc}
\tabletypesize{\scriptsize}
\tablecolumns{9}
\tablewidth{0pc}
\tablecaption{Members of B2319 Foreground Group}
\tablehead{
\colhead{Label} &
\colhead{RA} &
\colhead{Dec} &
\colhead{$z$} &
\colhead{B\tablenotemark{a}} &
\colhead{V\tablenotemark{a}} &
\colhead{I\tablenotemark{a}} &
\colhead{Morphology\tablenotemark{b}} &
\colhead{Emission} 
}
\startdata
2319\_1\_1 & 23 21 37.25 & 05 25 09.2 & 0.5867 & 24.28 & 23.61 & 21.85 & \nodata & Yes\\
2319\_1\_2 & 23 21 43.37 & 05 25 43.9 & 0.5868 & 24.47 & 23.01 & 20.70 & \nodata & No\\
2319\_1\_3 & 23 21 37.54 & 05 26 22.1 & 0.5893 & 23.27 & 22.39 & 20.39 & Irr & Yes\\
2319\_1\_4 & 23 21 38.24 & 05 27 11.3 & 0.5861 & 23.96 & 22.09 & 19.59 & E & No\\
2319\_1\_5 & 23 21 39.39 & 05 27 19.2 & 0.5917 & 26.12 & 23.94 & 21.51 & E & Yes\\
2319\_1\_6 & 23 21 40.73 & 05 27 23.9 & 0.5922 & 24.24 & 23.94 & 22.75 & Irr & Yes\\
2319\_1\_7 & 23 21 38.23 & 05 27 38.3 & 0.5913 & 23.27 & 21.70 & 19.11 & E & No\\
2319\_1\_8 & 23 21 38.41 & 05 27 33.2 & 0.5876 & 24.97 & 23.12 & 20.45 & E & No\\
2319\_1\_9 & 23 21 38.82 & 05 27 31.5 & 0.5910 & 25.73 & 24.47 & 22.31 & Sa & No\\
2319\_1\_10 & 23 21 40.61 & 05 27 38.2 & 0.5888 & 25.08 & 23.55 & 21.11 & Irr & Yes\\
\enddata
\tablenotetext{a}{From Keck LRIS imaging}
\tablenotetext{b}{Determined from WFPC2 F555W and F814W imaging when available}
\label{table_2319fore}
\end{deluxetable*}

\begin{deluxetable*}{lccclllcc}
\tabletypesize{\scriptsize}
\tablecolumns{9}
\tablewidth{0pc}
\tablecaption{Members of B2319 Background Group}
\tablehead{
\colhead{Label} &
\colhead{RA} &
\colhead{Dec} &
\colhead{$z$} &
\colhead{B\tablenotemark{a}} &
\colhead{R\tablenotemark{a}} &
\colhead{I\tablenotemark{a}} &
\colhead{Morphology\tablenotemark{b}} &
\colhead{Emission} 
}
\startdata
2319\_2\_1 & 23 21 37.17 & 05 25 57.2 & 0.6887 & 25.95 & 24.72 & 22.13 & \nodata & No\\
2319\_2\_2 & 23 21 39.98 & 05 26 07.8 & 0.6891 & 23.97 & 23.69 & 22.65 & \nodata & Yes\\
2319\_2\_3 & 23 21 40.65 & 05 26 27.7 & 0.6905 & 25.31 & 23.36 & 20.93 & \nodata & No\\
2319\_2\_4 & 23 21 41.99 & 05 26 40.0 & 0.6881 & 23.79\tablenotemark{c} & 22.13\tablenotemark{c} & 19.31\tablenotemark{c} & \nodata & No\\
2319\_2\_5 & 23 21 41.99 & 05 26 40.0 & 0.6884 & \nodata\tablenotemark{c} & \nodata\tablenotemark{c} & \nodata\tablenotemark{c} & \nodata & No\\
2319\_2\_6 & 23 21 43.02 & 05 26 48.7 & 0.6879 & 25.87 & 24.43 & 21.71 & \nodata & No\\
2319\_2\_7 & 23 21 41.97 & 05 26 56.4 & 0.6903 & 25.05 & 24.45 & 22.82 & \nodata & Yes\\
2319\_2\_8 & 23 21 39.16 & 05 28 29.3 & 0.6898 & 24.08 & 23.55 & 21.92 & S0 & Yes\\
\enddata
\tablenotetext{a}{From Keck LRIS imaging}
\tablenotetext{b}{Determined from WFPC2 F555W and F814W imaging when available}
\tablenotetext{c}{We are not able to extract individual magnitudes for these objects from our imaging data; however, the objects have distinct spectral traces, so we report them as individual galaxies}
\label{table_2319back}
\end{deluxetable*}

The groups associated with B2319 have more varied populations than the B1600 group (see Tables \ref{table_2319fore} and \ref{table_2319back}). Using our characterization scheme, the early-type fractions for these groups are 0.5 and 0.63 for the foreground and background groups. These early-type fractions are typical of groups in the local Universe that tend to also have X-ray emission associated with them \citep{zabludoff,mulchaey}, although B2319 has not yet been observed at X-ray wavelengths. These early-type fractions lead us to believe that these two groups are likely to be bound and relaxed structures, and therefore the mass estimates derived from the measured velocity dispersions approximate the masses of these groups. 

\begin{figure*}
\begin{center}
 \includegraphics[width=0.3\textwidth]{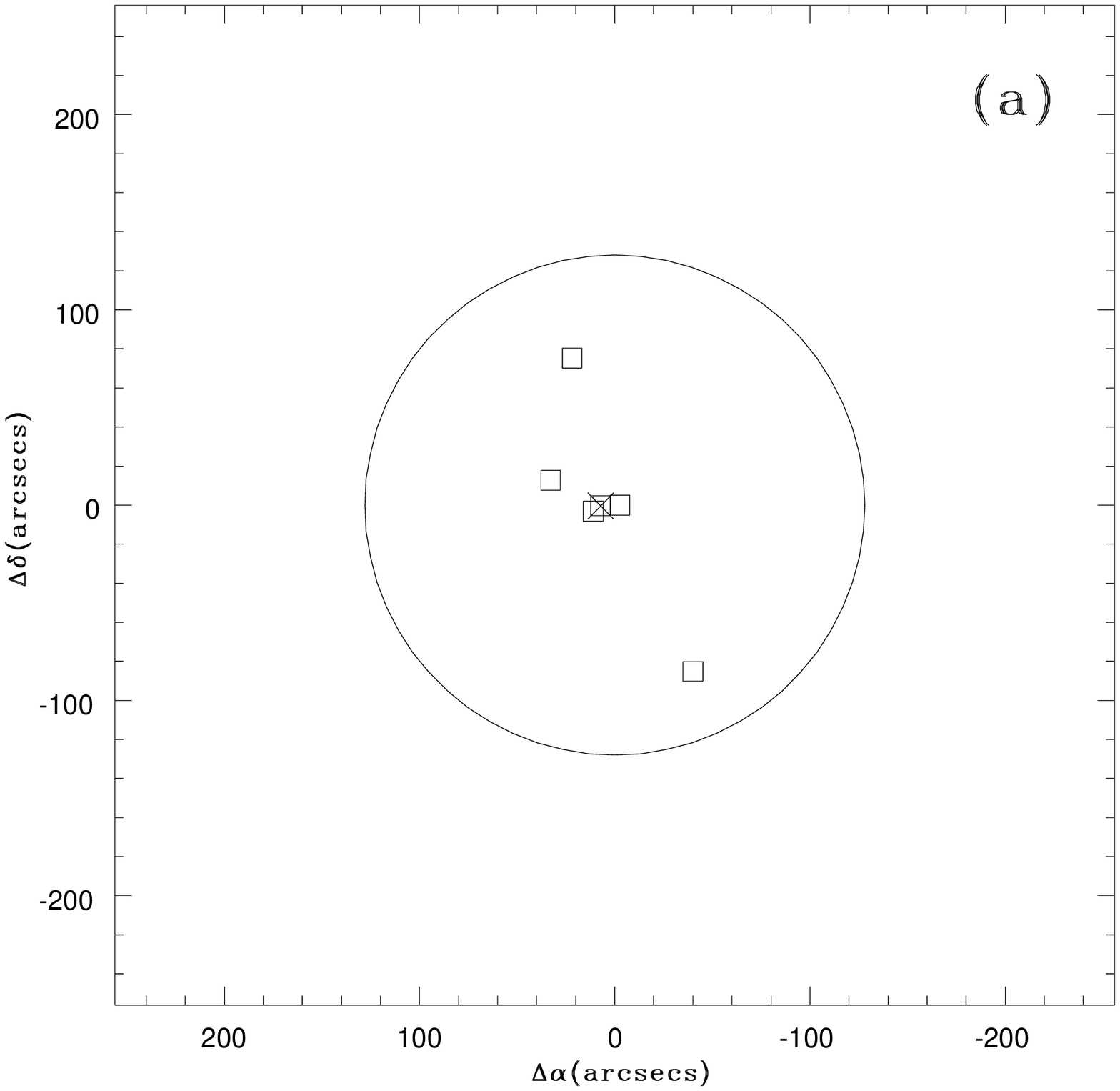}
 \includegraphics[width=0.3\textwidth]{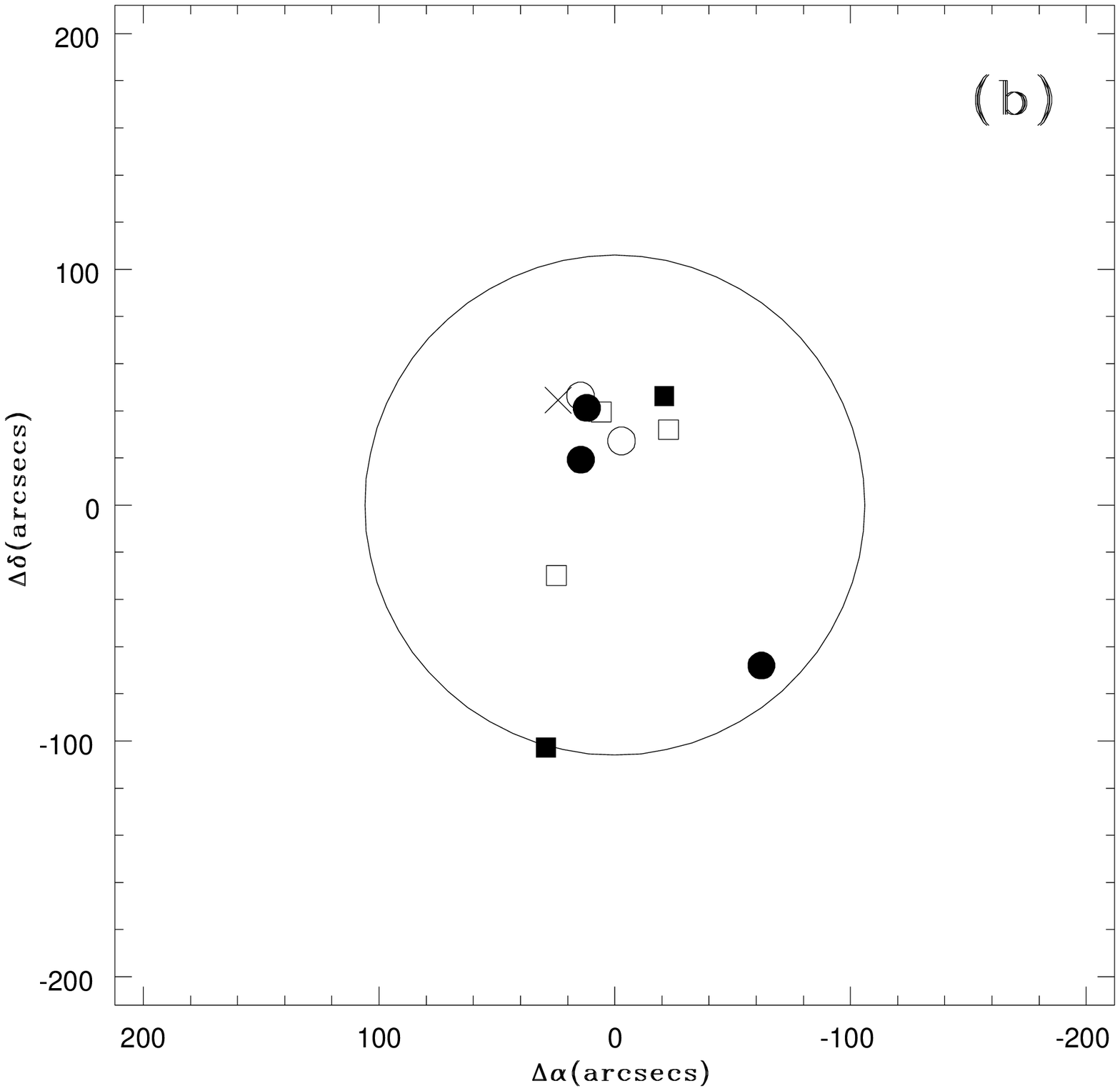}
 \includegraphics[width=0.3\textwidth]{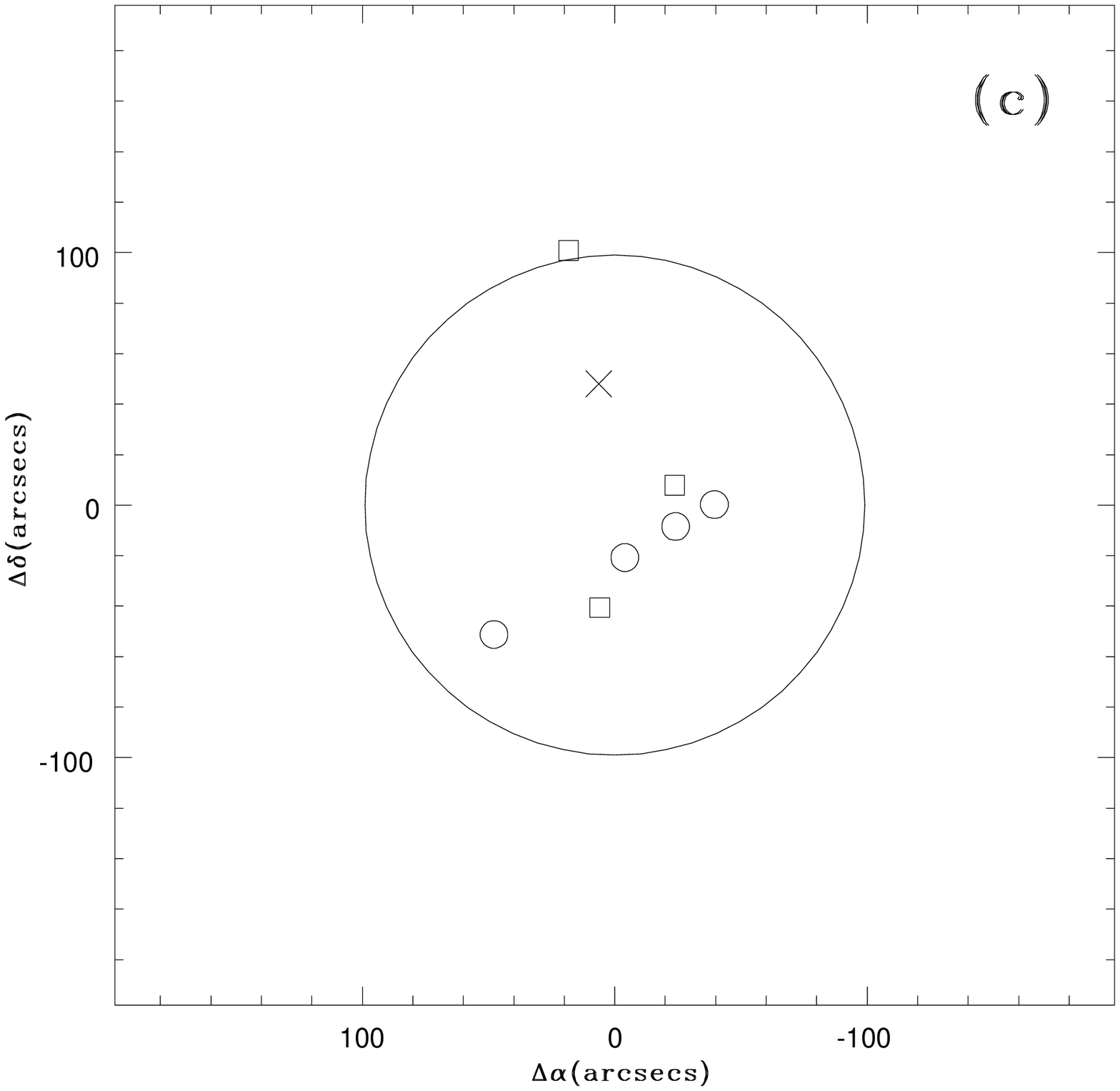}
  \caption{Spatial distribution of early-type (circles) and late-type (squares) galaxies for the (a) B1600 group and the (b) foreground and (c) background groups of B2319. The fields are centered on the group center and are 2~$h^{-1}$Mpc on a side. The central circle has a radius of 0.5~$h^{-1}$Mpc, and the X marks the position of the lensing galaxy. Late-type galaxies are marked with squares and early-types with circles. The shaded markers in (b) represent galaxies with negative velocities in the bimodal velocity distribution of Figure \ref{velocity}b and the open markers are galaxies with positive velocities.}
\label{fig_positions}
\end{center}
\end{figure*}

In Figure \ref{fig_positions} we plot the locations of the early-type (circles) and late-type (squares) galaxies in each group with respect to the group center.  Each field is 2~$h^{-1}$Mpc on a side and the central circle has a radius of 0.5~$h^{-1}$Mpc. We also show the location of the lensing galaxy, denoted by an X, with respect to the group centers. For the B2319 foreground group, we mark the galaxies with positive velocities (see Figure \ref{velocity}b) with open symbols and galaxies with negative velocities with filled symbols.

The center of the B1600 group is coincident with the lensing galaxy, although the overall structure of the group is somewhat filamentary.  This elongated structure could be due to only having one slitmask for the group, causing a bias in the selection of target galaxies; the group is roughly aligned along the North-South axis, which is also the orientation axis of the slitmask. The distribution of the inner four galaxies of the group is very compact, and the group meets all of the criteria of \citet{hickson} for compact groups except the requirement that the group be isolated from other galaxies (to distinguish compact groups from cluster--or in this case group--cores, for example).

We find that neither of the groups associated with B2319 are centered on the lensing galaxy. The background group has an elongated structure similar to the B1600 group, although this B2319 group is not preferentially aligned with the slitmasks used for the field. The B2319 groups do not show a preference for early-type galaxies lying in the center of the groups, contrary to what one might expect for relaxed groups. However, the foreground group appears to be quite compact with seven members lying in a region with a radius of $\sim 100 h^{-1}$~kpc. Additionally, there is not a clear separation between the positive velocity galaxies and the negative velocity galaxies (see Figure \ref{velocity}b), indicating that the bimodality of the velocity distribution for the foreground group might be a result of incomplete sampling.

\section{LENS MODELS}
There are several ways to apply the group environments to the lensing models of B2319 and B1600.  One method is to treat the group as a single dark matter halo and model the group as a smooth mass distribution.  Alternatively, we could model each of the group members individually.  This allows the possibility that our groups might be chance associations of galaxies instead of bound systems.  Finally, we may model the group as lumpy masses embedded in a smooth distribution, a combination of the former two cases.  This model of a halo with substructure is consistent with cluster observations and theoretical predictions \citep{natarajan,gao,zentner}. However, due to incomplete information about our groups, in this analysis we only use the first two methods to give approximate indications for the contributions of the group environments to the lenses. For simplicity, we model all halos as singular isothermal spheres (SIS).

\subsection{B2319 Lens Models}
Models for B2319 were found to inadequately reproduce the image configuration without applying a large shear term \citep[$\gamma = 0.14, \rm{PA}_{\gamma} = -22^{\circ}$;][]{rusin}.  The necessary additional shear is essentially perpendicular to the expected orientation of the shear caused by the lensing galaxy, and this excess shear is therefore assumed to be caused by another object. We investigate whether the foreground and background groups we have detected can adequately account for this discrepancy. Although the groups associated with B2319 are not at the same redshift as the primary lensing galaxy, we calculate the shear at the lensing plane of the groups and identify this as the same shear required by \citet{rusin} \citep[see also][]{fassnacht2005,momcheva}.

We create two models for the system by accounting for the foreground and background groups and for the individual galaxies of both groups. The groups and the individual galaxies are all modeled with SIS profiles, where the brightest galaxy in each group is assigned a fiducial velocity dispersion of $\sigma = 200 {\rm km}\,{\rm s}^{-1}$ and the contributions of the other galaxies are scaled by $10^{-0.2(m - m_{fid})}$ where $m_{fid}$ is the magnitude of the brightest galaxy \citep[e.g.,][]{keeton2004,fassnacht2005}. For the group model, we use the luminosity-weighted average position of the group galaxies to determine the group centroid. The results for these models are collected in Table \ref{table_2319_models}. We find a total additional shear of $\gamma = 0.05$ at a PA of 33$^\circ$ for the group halo model and a shear of $\gamma = 0.023$ at a PA of 86$^\circ$ for the individual galaxies model. This correction differs substantially from the shear required by \citet{rusin} in both magnitude and orientation, although this could be caused by our uncertainty in the true position of the group centroid.

\begin{deluxetable*}{lccccccccc}
\tabletypesize{\scriptsize}
\tablecolumns{10}
\tablewidth{0pc}
\tablecaption{B2319 Lens Environment Modeling}
\tablehead{
 \colhead{Model} &
 \colhead{$\kappa_{frg}$} &
 \colhead{$\gamma_{frg}$} &
 \colhead{PA$_{frg}$} &
 \colhead{$\kappa_{bkg}$} &
 \colhead{$\gamma_{bkg}$} &
 \colhead{PA$_{bkg}$} &
 \colhead{$\kappa_{total}$} &
 \colhead{$\gamma_{total}$} &
 \colhead{PA$_{total}$}
}
\startdata
Groups & 0.046 & 0.046 & 37 & 0.006 & 0.006 & -7 & 0.051 & 0.050 & 33\\
Individual Galaxies & 0.033 & 0.025 & 100 & 0.008 & 0.006 & -14 & 0.041 & 0.023 & 86\\
\enddata
\label{table_2319_models}
\tablecomments{Values marked \emph{frg} are for the foreground group, those marked \emph{bkg} are for the background group, and those marked \emph{total} are for the sum of the contributions of the foreground and background groups.}
\end{deluxetable*}

\subsection{B1600 Lens Models}
The model for B1600+434 adequately reproduces the observed image distribution with physically reasonable parameters \citep{koopmans1998}. However, the presence of the group that includes the lensing galaxy adds an additional mass sheet to the lens plane. This mass sheet increases the convergence in the lens model, which in turn effects the value of $H_0$ determined for this lens system. The additional convergence will decrease the value of $H_0$ compared to models that do not account for the lens environment:
$$
H_{0,true} = H_{0,meas}~(1 - \kappa_{group}).
$$
In this analysis, we calculate the expected added convergence due to the group, due to the individual galaxies excluding the lensing galaxy, and due to the individual galaxies excluding the lens and its neighbor galaxy. The neighbor galaxy has already been included in some lens models \citep[see][for example]{maller}. We use the luminosity-weighted positions of the group members to determine the group centroid for the group SIS. Results for each of the models are collected in Table \ref{table_1600_model}. Previous analyses of B1600 have derived values of $H_0$ between $52~{\rm km}\,{\rm s}^{-1}\,{\rm Mpc}^{-1}$ and $60~{\rm km}\,{\rm s}^{-1}\,{\rm Mpc}^{-1}$ \citep{burud,koopmans2000}, already lower than values determined by other methods \citep[e.g.,][]{sanchez,tegmark,freedman}. The additional convergence due to the environment of B1600 lowers these values by a further $\sim 5\%$.

\begin{deluxetable*}{lccc}
\tabletypesize{\scriptsize}
\tablecolumns{4}
\tablewidth{0pc}
\tablecaption{B1600 Lens Environment Modeling}
\tablehead{
 \colhead{Model} &
 \colhead{$\kappa_{env}$} &
 \colhead{$\gamma_{env}$} &
 \colhead{PA$_{env}$}
}
\startdata
Group & 0.014 & 0.014 & 18 \\
Individual galaxies\tablenotemark{a} & 0.050 & 0.031 & -53 \\
Individual galaxies\tablenotemark{b} & 0.012 & 0.008 & 98 \\
\enddata
\tablenotetext{a}{Excluding the lensing galaxy}
\tablenotetext{b}{Excluding the lensing galaxy and its neighbor}
\label{table_1600_model}
\end{deluxetable*}

\section{DISCUSSION AND CONCLUSIONS}
The absence of early-type galaxies and X-ray emission indicates that the B1600 group may not be a relaxed or bound system. Furthermore, the elongated structure of the group may favor this unrelaxed interpretation, although groups have been shown to have a wide range of morphologies \citep{zabludoff,wilmana}. Though our sampling is not complete enough to make strong statements regarding the characteristic length and mass scales for this group, we find the mass to be consistent with other groups with similar membership \citep[e.g.,][]{zabludoff,ramella}. However, the arguments for and against the bound interpretation of these galaxies all rely on comparisons with local groups. It seems perfectly reasonable to expect that groups at higher redshifts are dynamically younger and therefore their constituent galaxies have not had time to undergo processing. This would account for the absence of early-type galaxies and hot intragroup gas. The compactness of the distribution of the central members of the group and the presence of a strong gravitational lens are indicators that, at a minimum, the \emph{central structure} of the group is bound in a common halo, and we therefore conclude that B1600 lies in a group environment.

The groups associated with B2319 are similar to groups found in the local Universe; the numbers of members, early-type fractions, and velocity dispersions of these two groups are typical of local groups. The foreground group is potentially a poor cluster, and is a good candidate for X-ray followup \citep[e.g.,][]{jeltema}. The nature of the background group is more difficult to discern; the group has an irregular morphology with no proper core but a large early-type fraction. The absence of a core might portend a group that is forming, though the large early-type fraction suggests the constituent galaxies have had several significant interactions. It is worth noting that we did not target galaxies at a redshift of $z \sim 0.7$, and more complete spectroscopy of the field might make a core structure more apparent for the group. In any case, the number of members (8) and the approximately gaussian nature of the velocity histogram for this group (Figure \ref{velocity}c) give us confidence that this is a group associated with a common halo.

Accounting for the environments of gravitational lenses in the model of the potential is theoretically straightforward but quite difficult to do in practice. The lens corrections based upon individual galaxies rely on incomplete membership information and incorrect modeling of velocity dispersions for individual group members. These problems are probably less severe than those present for the group halo model. Velocity dispersion estimates for group halos can be off by significant amounts because of incomplete group membership sampling \citep{zabludoff}. Due to the quadratic dependence of SIS models on velocity dispersions, modeling errors for SIS models can be substantial for poorly sampled groups. Another important source of error is the modeling of centroids for the groups. Different centroiding methods can yield very different angular offsets of the group from the lens due to sampling small numbers of the group members \citep[e.g.,][]{fassnacht2002,fassnacht2005}. Additionally, the SIS model assumes a relaxed group halo, though this might not be the case for many moderate redshift groups (see the discussion of the B1600 group above).

Deep X-ray observations of lens systems would potentially correct for all of the problems associated with modeling group halos; the presence of diffuse X-ray emission would indicate the presence of a group halo, the X-ray temperature could be used as a mass measure, and the diffuse X-ray centroid would presumably coincide with the mass centroid \citep{mulchaey}. However, for all but the most massive groups, it is very difficult to detect intragroup X-ray emission due to the low luminosities and cosmological dimming of these moderate redshift sources (Dai \& Kochanek 2005; Neureuther et al. 2006, \emph{in prep}), although some detections have been made \citep{jeltema,mulchaey2006,grant}. Thus, for our groups with only optical information, we favor using the individual galaxies to interpret our results as lower bounds on lens corrections due to group environments.

We find that the environments for each lens system contribute approximately 5\% to the convergence in the lensing models, as shown in Tables \ref{table_2319_models} and \ref{table_1600_model}, similar to results obtained by \citet{momcheva}. For B1600, this suggests a commensurate 5\% decrease in the determined value of $H_0$. Note that ignoring lens environments causes a \emph{systematically} inflated value of $H_0$, and our analysis simply removes (some of) this bias; this should be considered a reinterpretation of the previously published values of $H_0$ for this lens system and not a new calculation for the Hubble Constant. We also find that the environment adds a small contribution to the shear, although we emphasize that this is \emph{very} dependent on the membership determined for the groups. However, we targeted most reasonable candidates within $\sim 10\arcsec$ of the lenses, leading us to believe that we have complete membership information for galaxies that would most significantly affect the lensing. We are in the process of investigating several more lens systems and we will attempt to quantify uncertainties in our convergence and shear estimates based upon this larger sample.


\acknowledgments 
Based in part on observations made with the NASA/ESA Hubble Space Telescope, obtained from the the Data Archive at the Space Telescope Science Institute (STScI). STScI is operated by the Association of Universities for Research in Astronomy, Inc., under NASA contract NAS5-26555. These observations are associated with program \#AR-10300, supported by NASA through a grant from STScI. Additionally, some of the data presented herein were obtained at the W.M. Keck Observatory, which is operated as a scientific partnership among the California Institute of Technology, the University of California and the National Aeronautics and Space Administration. The Observatory was made possible by the generous financial support of the W.M. Keck Foundation. The authors wish to recognize and acknowledge the very significant cultural role and reverence that the summit of Mauna Kea has always had within the indigenous Hawaiian community.  We are most fortunate to have the opportunity to conduct observations from this mountain. This work is also based in part on data collected at the Subaru Telescope and obtained from the SMOKA science archive at the Astronomical Data Analysis Center, which are operated by the National Astronomical Observatory of Japan.



\begin{thebibliography}{}

\bibitem[Aarseth \& Fall(1980)]{aarseth} Aarseth, S.~J., \& 
Fall, S.~M.\ 1980, \apj, 236, 43

\bibitem[Abazajian et al.(2005)]{abazajian} Abazajian, K., et 
al.\ 2005, \apj, 625, 613 

\bibitem[Aceves \& Vel{\' a}zquez(2002)]{aceves} Aceves, H., 
\& Vel{\' a}zquez, H.\ 2002, Revista Mexicana de Astronomia y Astrofisica, 
38, 199 

\bibitem[Baba et al.(2002)]{baba} Baba, H., et al.\ 2002, 
ASP Conf.~Ser.~281: Astronomical Data Analysis Software and Systems XI, 
281, 298

\bibitem[Bahcall et al.(1997)]{bahcall} Bahcall, J.~N., 
Kirhakos, S., Saxe, D.~H., \& Schneider, D.~P.\ 1997, \apj, 479, 642 

\bibitem[Balogh et al.(2004)]{balogh} Balogh, M., et al.\ 
2004, \mnras, 348, 1355

\bibitem[Best(2004)]{best2004} Best, P.~N.\ 2004, \mnras, 351, 
70

\bibitem[Best et al.(2005)]{best2005} Best, P.~N., Kauffmann, 
G., Heckman, T.~M., Brinchmann, J., Charlot, S., Ivezi{\' c}, {\v Z}., \& 
White, S.~D.~M.\ 2005, \mnras, 660 

\bibitem[Blandford et al.(2001)]{blandford} Blandford, R., Surpi, 
G., \& Kundi{\' c}, T.\ 2001, ASP Conf.~Ser.~237: Gravitational Lensing: 
Recent Progress and Future Goals, 237, 65 

\bibitem[Bower \& Balogh(2004)]{bower} Bower, R.~G., \& 
Balogh, M.~L.\ 2004, Clusters of Galaxies: Probes of Cosmological Structure 
and Galaxy Evolution, 326

\bibitem[Browne et al.(2003)]{browne} Browne, I.~W.~A., et 
al.\ 2003, \mnras, 341, 13 

\bibitem[Burud et al.(2000)]{burud} Burud, I., et al.\ 2000, 
\apj, 544, 117

\bibitem[Carlberg et al.(2001a)]{carlberga} Carlberg, R.~G., Yee, 
H.~K.~C., Morris, S.~L., Lin, H., Hall, P.~B., Patton, D.~R., Sawicki, M., 
\& Shepherd, C.~W.\ 2001, \apj, 552, 427 

\bibitem[Carlberg et al.(2001b)]{carlbergb} Carlberg, R.~G., Yee, 
H.~K.~C., Morris, S.~L., Lin, H., Hall, P.~B., Patton, D.~R., Sawicki, M., 
\& Shepherd, C.~W.\ 2001, \apj, 563, 736

\bibitem[Chae et al.(2001)]{chae} Chae, K.-H., Mao, S., \& 
Augusto, P.\ 2001, \mnras, 326, 1015

\bibitem[Coil et al.(2004)]{coil2004} Coil, A.~L., et al.\ 2004, 
\apj, 609, 525

\bibitem[Coil et al.(2006)]{coil2006} Coil, A.~L., et al.\ 2006, 
\apj, 638, 668 

\bibitem[Colless et al.(2001)]{colless} Colless, M., et al.\ 
2001, \mnras, 328, 1039

\bibitem[Collister \& Lahav(2005)]{collister} Collister, A.~A., 
\& Lahav, O.\ 2005, \mnras, 361, 415 

\bibitem[Conselice et al.(2003)]{conselice} Conselice, C.~J., 
Bershady, M.~A., Dickinson, M., \& Papovich, C.\ 2003, \aj, 126, 1183

\bibitem[Dai \& Kochanek(2005)]{dai} Dai, X., \& Kochanek, 
C.~S.\ 2005, \apj, 625, 633 

\bibitem[Dalal \& Watson(2004)]{dalal} Dalal, N., 
\& Watson, C.~R.\ 2004, ArXiv Astrophysics e-prints, arXiv:astro-ph/0409483

\bibitem[D'Onghia \& Lake(2004)]{donghia} D'Onghia, E., \& 
Lake, G.\ 2004, \apj, 612, 628

\bibitem[Evrard et al.(2002)]{evrard} Evrard, A.~E., et al.\ 
2002, \apj, 573, 7

\bibitem[Fassnacht \& Cohen(1998)]{fassnacht1998} Fassnacht, C.~D., 
\& Cohen, J.~G.\ 1998, \aj, 115, 377

\bibitem[Fassnacht \& Lubin(2002)]{fassnacht2002} Fassnacht, C.~D., 
\& Lubin, L.~M.\ 2002, \aj, 123, 627 

\bibitem[Fassnacht et al.(2005)]{fassnacht2005} Fassnacht, C.~D., 
Gal, R.~R., Lubin, L.~M., McKean, J.~P., Squires, G.~K., \& Readhead, 
A.~C.~S.\ 2006, \apj, in press (astro-ph/0510728)

\bibitem[Faure et al.(2004)]{faure} Faure, C., Alloin, D., 
Kneib, J.~P., \& Courbin, F.\ 2004, \aap, 428, 741

\bibitem[Freedman et al.(2001)]{freedman} Freedman, W.~L., et 
al.\ 2001, \apj, 553, 47

\bibitem[Gao et al.(2004)]{gao} Gao, L., White, S.~D.~M., 
Jenkins, A., Stoehr, F., \& Springel, V.\ 2004, \mnras, 355, 819 

\bibitem[Gerke et al.(2005)]{gerke} Gerke, B.~F., et al.\ 
2005, \apj, 625, 6

\bibitem[G{\' o}mez et al.(2003)]{gomez} G{\' o}mez, P.~L., 
et al.\ 2003, \apj, 584, 210

\bibitem[Grant et al.(2004)]{grant} Grant, C.~E., Bautz, 
M.~W., Chartas, G., \& Garmire, G.~P.\ 2004, \apj, 610, 686

\bibitem[Hickson(1982)]{hickson} Hickson, P.\ 1982, \apj, 255, 
382

\bibitem[Hoekstra et al.(2001)]{hoekstra} Hoekstra, H., et al.\ 
2001, \apjl, 548, L5

\bibitem[Jackson et al.(1995)]{jackson} Jackson, N., et al.\ 
1995, \mnras, 274, L25 

\bibitem[Jeltema et al.(2006)]{jeltema} Jeltema, T.~E.,
Mulchaey, J.~S., Lubin, L.~M., Rosati, P., B{\" o}hringer, H.\ 2006,
\apj, submitted

\bibitem[Kauffmann et al.(2004)]{kauffmann} Kauffmann, G., White, 
S.~D.~M., Heckman, T.~M., M{\' e}nard, B., Brinchmann, J., Charlot, S., 
Tremonti, C., \& Brinkmann, J.\ 2004, \mnras, 353, 713 

\bibitem[Keeton et al.(2000)]{keeton2000} Keeton, C.~R., 
Christlein, D., \& Zabludoff, A.~I.\ 2000, \apj, 545, 129 

\bibitem[Keeton \& Zabludoff(2004)]{keeton2004} Keeton, C.~R., \& 
Zabludoff, A.~I.\ 2004, \apj, 612, 660 

\bibitem[Kochanek(2002)]{kochanek2002} Kochanek, C.~S.\ 2002, \apj, 
578, 25 

\bibitem[Kochanek(2003)]{kochanek2003} Kochanek, C.~S.\ 2003, \apj, 
583, 49

\bibitem[Kochanek \& Schechter(2004)]{kochanek2004} Kochanek, C.~S., 
\& Schechter, P.~L.\ 2004, Measuring and Modeling the Universe, 117

\bibitem[Koopmans et al.(1998)]{koopmans1998} Koopmans, L.~V.~E., de 
Bruyn, A.~G., \& Jackson, N.\ 1998, \mnras, 295, 534 

\bibitem[Koopmans et al.(2000)]{koopmans2000} Koopmans, L.~V.~E., de 
Bruyn, A.~G., Xanthopoulos, E., \& Fassnacht, C.~D.\ 2000, \aap, 356, 391

\bibitem[Koopmans et al.(2003)]{koopmans2003} Koopmans, L.~V.~E., 
Treu, T., Fassnacht, C.~D., Blandford, R.~D., \& Surpi, G.\ 2003, \apj, 
599, 70 

\bibitem[Leh{\'a}r et al.(2000)]{lehar} Leh{\'a}r, J., et 
al.\ 2000, \apj, 536, 584 

\bibitem[Lin et al.(2004)]{lin} Lin, L., et al.\ 2004, 
\apjl, 617, L9

\bibitem[Lubin et al.(2000)]{lubin} Lubin, L.~M., Fassnacht, 
C.~D., Readhead, A.~C.~S., Blandford, R.~D., \& Kundi{\' c}, T.\ 2000, \aj, 
119, 451

\bibitem[Maller et al.(2000)]{maller} Maller, A.~H., Simard, 
L., Guhathakurta, P., Hjorth, J., Jaunsen, A.~O., Flores, R.~A., \& 
Primack, J.~R.\ 2000, \apj, 533, 194

\bibitem[McLure \& Dunlop(2001)]{mclure} McLure, R.~J., \& 
Dunlop, J.~S.\ 2001, \mnras, 321, 515 

\bibitem[Merch{\'a}n \& Zandivarez(2005)]{merchan} Merch{\'a}n, 
M.~E., \& Zandivarez, A.\ 2005, \apj, 630, 759 

\bibitem[Miller et al.(2003)]{miller} Miller, C.~J., Nichol, 
R.~C., G{\' o}mez, P.~L., Hopkins, A.~M., \& Bernardi, M.\ 2003, \apj, 597, 
142 

\bibitem[Miyazaki et al.(2002)]{miyazaki} Miyazaki, S., et al.\ 
2002, \pasj, 54, 833

\bibitem[M{\" o}ller et al.(2002)]{moller} M{\" o}ller, O., 
Natarajan, P., Kneib, J.-P., \& Blain, A.~W.\ 2002, \apj, 573, 562 

\bibitem[Momcheva et al.(2005)]{momcheva} Momcheva, I., 
Williams, K.~A., Keeton, C.~R., \& Zabludoff, A.~I.\ 2005, ArXiv 
Astrophysics e-prints, arXiv:astro-ph/0511594

\bibitem[Morgan et al.(2005)]{morgan} Morgan, N.~D., Kochanek, 
C.~S., Pevunova, O., \& Schechter, P.~L.\ 2005, \aj, 129, 2531

\bibitem[Mulchaey et al.(2003)]{mulchaey} Mulchaey, J.~S., 
Davis, D.~S., Mushotzky, R.~F., \& Burstein, D.\ 2003, \apjs, 145, 39

\bibitem[Mulchaey et al.(2006)]{mulchaey2006} Mulchaey, J.~S.,
Lubin, L.~M., Fassnacht, C.~D., Rosati, P., \& Jeltema, T.~E.\ 2006,
\apj, submitted

\bibitem[Myers et al.(2003)]{myers} Myers, S.~T., et al.\ 
2003, \mnras, 341, 1 

\bibitem[Natarajan \& Springel(2004)]{natarajan} Natarajan, P., 
\& Springel, V.\ 2004, \apjl, 617, L13 

\bibitem[Oguri et al.(2005)]{oguri} Oguri, M., Keeton, C.~R., 
\& Dalal, N.\ 2005, \mnras, 364, 1451 

\bibitem[Ouchi et al.(2004)]{ouchi} Ouchi, M., et al.\ 2004,
\apj, 611, 660

\bibitem[Oke et al.(1995)]{oke} Oke, J.~B., et al.\ 1995, 
\pasp, 107, 375 

\bibitem[Padilla et al.(2004)]{padilla} Padilla, N.~D., et al.\ 
2004, \mnras, 352, 211

\bibitem[Patton et al.(2002)]{patton} Patton, D.~R., et al.\ 
2002, \apj, 565, 208

\bibitem[Ramella et al.(1989)]{ramella} Ramella, M., Geller, 
M.~J., \& Huchra, J.~P.\ 1989, \apj, 344, 57

\bibitem[Reed et al.(2005)]{reed} Reed, D., Governato, F., 
Quinn, T., Gardner, J., Stadel, J., \& Lake, G.\ 2005, \mnras, 359, 1537

\bibitem[Rusin et al.(2001)]{rusin} Rusin, D., et al.\ 2001, 
\aj, 122, 591

\bibitem[S{\'a}nchez et al.(2006)]{sanchez} S{\'a}nchez, A.~G., 
Baugh, C.~M., Percival, W.~J., Peacock, J.~A., Padilla, N.~D., Cole, S., 
Frenk, C.~S., \& Norberg, P.\ 2006, \mnras, 366, 189 

\bibitem[Sheinis et al.(2002)]{sheinis} Sheinis, A.~I., Bolte, 
M., Epps, H.~W., Kibrick, R.~I., Miller, J.~S., Radovan, M.~V., Bigelow, 
B.~C., \& Sutin, B.~M.\ 2002, \pasp, 114, 851

\bibitem[Tanaka et al.(2004)]{tanaka} Tanaka, M., Goto, T., 
Okamura, S., Shimasaku, K., \& Brinkmann, J.\ 2004, \aj, 128, 2677

\bibitem[Tegmark et al.(2004)]{tegmark} Tegmark, M., et al.\ 
2004, \prd, 69, 103501

\bibitem[Tully(1987)]{tully} Tully, R.~B.\ 1987, \apj, 321, 
280

\bibitem[Weinmann et al.(2006)]{weinmann} Weinmann, S.~M., van 
den Bosch, F.~C., Yang, X., \& Mo, H.~J.\ 2006, \mnras, 366, 2

\bibitem[Williams et al.(2005)]{williams} Williams, K.~A., 
Momcheva, I., Keeton, C.~R., Zabludoff, A.~I., \& Lehar, J.\ 2005, ArXiv 
Astrophysics e-prints, arXiv:astro-ph/0511593 

\bibitem[Wilman et al.(2005a)]{wilmana} Wilman, D.~J., Balogh, 
M.~L., Bower, R.~G., Mulchaey, J.~S., Oemler, A., Carlberg, R.~G., Morris, 
S.~L., \& Whitaker, R.~J.\ 2005a, \mnras, 358, 71

\bibitem[Wilman et al.(2005b)]{wilmanb} Wilman, D.~J., et al.\ 
2005b, \mnras, 358, 88

\bibitem[Yang et al.(2004)]{yang} Yang, X., Mo, H.~J., Jing, 
Y.~P., van den Bosch, F.~C., \& Chu, Y.\ 2004, \mnras, 350, 1153

\bibitem[York et al.(2000)]{york} York, D.~G., et al.\ 2000, 
\aj, 120, 1579 

\bibitem[Zabludoff \& Mulchaey(1998)]{zabludoff} Zabludoff, 
A.~I., \& Mulchaey, J.~S.\ 1998, \apj, 496, 39 

\bibitem[Zentner et al.(2005)]{zentner} Zentner, A.~R., 
Berlind, A.~A., Bullock, J.~S., Kravtsov, A.~V., \& Wechsler, R.~H.\ 2005, 
\apj, 624, 505

\end{thebibliography}
\end{document}